\documentstyle[twoside,fleqn,espcrc2]{article}

\title{
The QCD analysis of the revised CCFR data for $xF_3$ structure
function: the next-to-next-to-leading order and Pad\'e approximants}
\author{A.L. Kataev \address{Institute for Nuclear Research
of the Academy of Sciences of Russia, 117312 Moscow, Russia},
A.V. Kotikov
\address{Particle Physics Laboratory, Joint
Institute for Nuclear Research, 141980 Dubna, Russia},
G. Parente \address{Department of Particle Physics, University
of Santiago de Compostela, 15706 Santiago de Compostela, Spain}
and A.V. Sidorov
\address{Bogoliubov Laboratory of Theoretical Physics, Joint
Institute for Nuclear Research, 141980 Dubna, Russia}}
\begin{document}
\begin{abstract}
The next-to-next-to-leading order (NNLO) QCD analysis of the
revised experimental data of the CCFR collaboration for the
$xF_3$ structure function of the deep-inelastic scattering
of neutrinos and antinuetrinos on the nucleons is made
by means of the Jacobi polynomial expansion technique.
The NNLO values
 of the QCD coupling constant are determined 
both without and with twist--4 contributions taken into account.
Theoretical ambiguities of $\alpha_s$
are fixed using the methods of the
Pad\'e approximants. We observe that the [0/2] Pad\'e approximant
is more appropriate for this purpose than the [1/1] one.

\end{abstract}
 
\maketitle
{\bf 1.}~
In the last several years the considerable
progress was achieved in the area of the analytical evaluation
of the characteristics of DIS  at  the
NNLO of perturbative QCD in the $\overline{MS}$-scheme.  Indeed,
the NNLO corrections were found  in Ref.\cite{VZ} for the
coefficient functions of the structure function (SF) $F_2$, while in
the works of Ref.\cite{VZ2} the similar NNLO calculations were made
for the SF $xF_3$. The results of Ref.\cite{VZ} are in agreement with
the expressions for the NNLO corrections to  the coefficient functions
of the non-singlet (NS) moments of $F_2$ SF, calculated
analytically in the case of the even moments with $n=2,4,6,8,10$ \cite{LV}.
Moreover, in the case of
$n=2,4,6,8,10$ the analytical expression of the NNLO corrections to
the anomalous dimensions of the NS moments of $F_2$ are  also
known at present \cite{LRV}. Together with the expression for the NNLO
correction to the QCD $\beta$-function \cite{Lesha}, the results of
Ref.\cite{VZ}-Ref.\cite{LRV}  are forming the
theoretical background for the extraction of the value of the
parameter $\Lambda_{\overline{MS}}^{(4)}$ from the NS fits of
the DIS data at the level of new theoretical precision, namely with
taking into account the effects of the NNLO corrections.

The classical approach of the incorporation of the information
about the QCD approximations of the SFs into the analysis
of the concrete DIS data is the DGLAP approach \cite{DGLAP}.
It is known that in its standard variant the DGLAP machinery
is based on the solution
of the corresponding integro-differential equations.
However, at the present level of the development of the calculation
technique it is still impossible to use this variant of the DGLAP
approach at the NNLO level.  Indeed, in spite of the fact that the
NNLO approximations of the DIS coefficient functions are already
known from the results of Refs.\cite{VZ,VZ2}, the explicit NNLO
expressions of the corresponding kernels are not yet calculated. This
still missed information can be obtained only after the
analytic calculations of the NNLO corrections to the anomalous
dimensions of the relevant composite operators in the case of
{\bf arbitrary} number of the moments $n$ will become available.

However, there are the methods, which are allowing to overcome this
limitation and to take into account the results of
the NNLO calculations
of Refs.\cite{VZ}-\cite{Lesha} in the process of the NNLO analysis
of the DIS data within the
framework of the Mellin image of the DGLAP equation.
  These methods are based on the application of the
information about the NNLO renormalization group evolution of the
{\bf finite} number of the Mellin moments of the DIS SFs and
on the further approximate reconstruction of the behavior of the
SFs themselves.
The first realization of the idea of
the reconstruction of the  SFs from the sets of the finite number of
their Mellin moments was proposed in Ref.\cite{Ynd}, where the series
in Bernstein polynomials was used. This idea was further generalized
in the works of
Ref.\cite{PS} where it was proposed to relate the SFs and
their Mellin moments using the series in the  Jacobi
polynomials. This method was developed in Ref.\cite{Jacoby}
and successfully applied in the process of the NLO fits of
the DIS data of
the BCDMS collaboration \cite{BCDMS} and of the old data of
the CCFR collaboration  of Ref.\cite{CCFR} (see
Ref.\cite{KatSid}).

The Jacobi polynomial technique was already used at the NNLO
in the process of the NS fits of the $F_2$ SF data
of the BCDMS collaboration \cite{PKK}, old $xF_3$ data
of the CCFR collaboration \cite{KKPS1} and the revised in
Ref.\cite{CCFR97} $xF_3$ data of  the CCFR collaboration
\cite{KKPS2}
(which we will denote hereafter as CCFR'97).

In this work we are summarizing the definite results of
Ref.\cite{KKPS2} and are generalizing them to the
next-to-next-to-next-to-leading order (N$^3$LO) level
by means of the Pad\'e
approximants approach.
Our aim is to get the estimate of the value of
$\Lambda^{(4)}_{\overline{MS}}$ and $\alpha_s(M_Z)$ at the
N$^3$LO and to fix more precisely
the theoretical uncertainties of the values of these parameters
, obtained in Ref.\cite{KKPS2}
at the NNLO level.

{\bf 2.}~
Let us  define the Mellin moments
for the NS SF $xF_3(x,Q^2)$:
$M_n^{NS}(Q^2)=\int_0^1 x^{n-1}F_3(x,Q^2)dx$
where $n=2,3,4,...$. The theoretical expression
for these moments
obey the following renormalization group equation
\begin{eqnarray}
\bigg(\mu\frac{\partial}{\partial\mu}+\beta(A_s)\frac{\partial}
{\partial A_s}
+\gamma_{NS}^{(n)}(A_s)\bigg)
\times \\ \nonumber
M_n^{NS}(Q^2/\mu^2,A_s(\mu^2))=0
\label{rg}
\end{eqnarray}
where $A_s=\alpha_s/(4\pi)$.
The renormalization group functions are defined as
$\mu\frac{\partial A_s}{\partial\mu}=\beta(A_s)=-2\sum_{i\geq 0}
\beta_i A_s^{i+2}$~~,
$\mu\frac{\partial ln
Z_n^{NS}}{\partial\mu}=\gamma_{NS}^{(n)}(A_s) =\sum_{i\geq 0}
\gamma_{NS}^{(i)}(n) A_s^{i+1}$
where
$ Z_n^{NS} $
are the renormalization constants of the corresponding
NS operators.  The solution of the renormalization group
equation
can be presented in the following form :
\begin{eqnarray}
\frac{M_n^{NS}(Q^2)}{M_n^{NS}(Q_0^2)}=
exp\bigg[-\int_{A_s(Q_0^2)}^{A_s(Q^2)}
\frac
{\gamma_{NS}^{(n)}(x)}{\beta(x)}dx\bigg]
\times \\ \nonumber
\frac{C_{NS}^{(n)}(A_s(Q^2))}
{C_{NS}^{(n)}(A_s(Q_0^2))}
\label{mom}
\end{eqnarray}
where $M_n^{NS}(Q_0^2)$ is the phenomenological quantity related to the
factorization scale dependent factor.
It can be parametrized through the parton distributions at fixed momentum
transfer $Q_0^2$ as
$M_n^{NS}(Q_0^2)=\int_0^1 x^{n-2} A(Q_0^2)x^{b(Q_0^2)}(1-x)^{c(Q_0^2)}
(1+\gamma(Q_0^2)x)dx$
with  $\gamma \neq 0$.

At the N$^3$LO the  expression for
the coefficient function $C_{NS}^{(n)}$  can be presented in the
following general form 
$C_{NS}^{(n)}(A_s)=1+C^{(1)}(n)A_s+C^{(2)}(n)A_s^2+C^{(3)}(n)A_s^3$,
while the corresponding expansion of the anomalous dimensions term
is
\begin{eqnarray}
exp\bigg[-\int^{A_s(Q^2)}
\frac{\gamma_{NS}^{(n)}(x)}{\beta(x)}dx\bigg]=&& \\ \nonumber
\big(A_s(Q^2)\big)^{\gamma_{NS}^{(0)}(n)/2\beta_0}
\times
[1+p(n)A_s(Q^2)          \\ \nonumber
+q(n)A_s(Q^2)^2+r(n)A_s(Q^2)^3]
\label{an}
\end{eqnarray}
where $C^{(1)}(n)$, $C^{(2)}(n)$, $p(n)$ and $q(n)$ are
defined in Ref. \cite{KKPS1}.

The coupling constant $A_s(Q^2)$ can be expressed
in terms
of the inverse powers of
  $L=\ln(Q^2/\Lambda_{\overline{MS}}^2)$
as
$A_s^{NNLO}=A_s^{NLO}+\Delta A_s^{NNLO}$ and
$A_s^{N^3LO}=A_s^{NNLO}+\Delta A_s^{N^3LO}$, where
\begin{equation}
A_s^{NLO}=\frac{1}{\beta_0
L}-\frac{\beta_1 ln(L)}{\beta_0^3 L^2}
\end{equation}
\begin{equation}
\Delta A_s^{NNLO}=\frac{1}{\beta_0^5 L^3}[\beta_1^2 ln^2 (L)
-\beta_1^2 ln(L) +\beta_2\beta_0-\beta_1^2]
\end{equation}
\begin{eqnarray}
\Delta A_s^{N^3LO}=\frac{1}{\beta_0^7 L^4}[\beta_1^3 (-ln^3 (L)
+\frac{5}{2}ln^2 (L)    \\ \nonumber
+2ln(L)-\frac{1}{2})-3\beta_1\beta_2 ln(L)
+\frac{\beta_3}{2}]~.
\end{eqnarray}

Notice that in our normalizations
 $\beta_0$, $\beta_1$, $\beta_2$ and $\beta_3$ read
$\beta_0=11-0.6667f$,
$\beta_1=102-12.6667f$,
$\beta_2=1428.50-279.611f+6.01852f^2$
\begin{table*}
\begin{tabular}{llllllll}
\multicolumn{8}{p{14cm}}{{\bf{Table 1.}}}         \\
\multicolumn{8}{p{14cm}}{The results of the QCD fit of
the CCFR'97 data for
$Q^2\geq 5 GeV^2$, $Q^2_0=5 GeV^2$ (86 experimental points).
Statistical errors are taken into account.}       \\
\hline
                & &        h(x)=0 &  &       & $ h(x)\neq 0  $  \\ \hline
                & &$ \chi^2$
& $\Lambda^{(4)}_{\overline{MS}}(MeV) $  &$\alpha_s(M^2_Z)   $
& $\chi^2$& $\Lambda^{(4)}_{\overline{MS}}(MeV)$  &$\alpha_s(M^2_Z)$  \\ \hline
LO &\cite{KKPS2}  & 113.3& 266$\pm$37&$0.135^{+0.003}_{-0.002}$  & 66.2   & 338$\pm$169 &$ 0.142^{+0.012}_{-0.017}$   \\
NLO &\cite{KKPS2} &  87.1& 341$\pm$41& 0.121$\pm$0.003 & 65.6   & 428$\pm$158 & $0.125^{+0.008}_{-0.010}$   \\
NNLO &\cite{KKPS2}&  78.4& 293$\pm$29& 0.119$\pm$0.002 & 65.7   & 264$\pm$85  & $0.117^{+0.006}_{-0.007} $  \\
N$^3$LO &Pad\'e    &  78.8& 307$\pm$31& 0.120$\pm$0.002 & 65.7  &  256$\pm$81& $0.117^{+.006}_{-0.007}  $   \\ \hline
\end{tabular}
\end{table*}
and
$\beta_3=29243.0-6946.30f+405.089f^2+1.49931f^3$
where the expression for $\beta_3$ was obtained in
Ref.\cite{4loop}. The inverse-log expansion for
$\Delta A_s^{N^3LO}$, which incorporates the information
about the coefficient $\beta_3$, was presented in Ref.\cite{ChKS}.

In order to generalize the considerations of  Refs.\cite{KKPS1,KKPS2}
to the level of the explicitly unknown N$^3$LO corrections we will
use the Pad\'e approximations approach of Refs.\cite{SEK,EGKS}.
This method
provides a possibility to estimate the higher order terms of
perturbation theory using the
Pad\'e approximants (in the case of the NLO analysis of
the experimental data
of the BCDMS and EMC collaborations
the similar idea was proposed in Ref.\cite{SP}).
In the framework of this
technique the values of $C^{(3)}(n)$  and $r(n)$  could be
expressed as

\begin{eqnarray}
Pade~[1/1]~: && C^{(3)}(n)=[C^{(2)}(n)]^2/C^{(1)}(n)        \\
&&   r(n)=q(n)^2/p(n) \\
Pade~[0/2]~: &&
C^{(3)}(n)=2C^{(1)}(n)C^{(2)}(n)    \\ \nonumber
&&  -[C^{(1)}(n)]^3       \\
&&   r(n)= 2p(n)q(n)-[p(n)]^3
\end{eqnarray}

Using the following equation
\begin{eqnarray}
&&xF_{3}^{N_{max}}(x,Q^2)=\frac{h(x)}{Q^2}+x^{\alpha}(1-x)^{\beta}
\times \\ \nonumber
&&\sum_{n=0}^{N_{max}}
\Theta_n ^{\alpha , \beta}
(x)\sum_{j=0}^{n}c_{j}^{(n)}{(\alpha ,\beta )}
M_{j+2,xF_3}       \left ( Q^{2}\right ),
\label{Jacobi}
\end{eqnarray}
one can reconstruct the SF from the corresponding Mellin moments.
Here $\Theta_n^{\alpha,\beta}$ are the Jacobi polynomials and
$\alpha,\beta$ are their parameters, fixed by the condition of the
requirement of the minimization of the error of the reconstruction of
the SF.  In our analysis
 we are considering the region $6\leq N_{max}\leq 10$.
The target mass corrections   are included in our  fits
up to order $O(Q^{-2})$-terms.
 In the same manner as in Refs.\cite{VM,KKPS2} we are  incorporating
in Eq.(11) the twist--4 contribution
$ h(x)/Q^2$
as the additional term.

{\bf 3}.~ The details of our NLO and NNLO fits, made
for the case of $f=4$ number of active flavours,
are described in Refs. \cite{KKPS1,KKPS2}. In order to perform
the N$^3$LO analysis we used the [0/2] Pad\'e approximant
motivated estimates of $C^{(3)}(n)$ and $r(n)$ terms (see
Eqs.(9),(10)). It should be stressed that the N$^3$LO [1/1] Pad\'e
approximant description of the CCFR'97 experimental data turned out
to be not acceptable, since  it produces rather high value
of $\chi^2$: $\chi^2/{nep}>2$ (where $nep=86$ is the number
of the experimental points taken into account). The similar
effect of the preference of the [0/2] Pad\'e approximant analysis
over the [1/1] one was found in Ref.\cite{EGKS} in the case
of the comparison
of the QCD
theoretical predictions for the polarized Bjorken sum rule (which
are closely related to the QCD predictions for the first
moment of the $xF_3$ SF, namely
for the Gross-Llewellyn Smith sum rule)
with the available experimental data.

The results for $\Lambda^{(4)}_{\overline{MS}}$, obtained at the LO,
NLO, NNLO and N$^3$LO, are presented in Table 1.
Using the LO, NLO, NNLO and N$^3$LO variants of the rigorous
$\overline{MS}$-scheme matching conditions, derived in Ref.\cite{ChKS}
following the lines of Ref.\cite{BW}, we are transforming
$\Lambda_{\overline{MS}}^{(4)}$ values
through the threshold of the production of the fifth flavour
$M_5=m_b$ (where $m_b$ is the $b$-quark pole mass) and are obtaining
the related values of $\Lambda^{(5)}_{\overline{MS}}$ and
$\alpha_s(M_Z)$.  

The results of the LO, NLO and NNLO extractions of the twist-4
parameter $h(x)$ from the CCFR'97 $xF_3$ data were presented in
Ref.\cite{KKPS2}. Using the Pad\'e approximations approach we
also found the $x$-form of $h(x)$ at the N$^3$LO. It turned out
to be almost undistinguished from the results of the NNLO fits.
Dew to the lack of space we are postponing the presentation of
the N$^3$LO results for $h(x)$ for the future
publication.

In the case of the non-zero values of the twist-4 parameter
$h(x)\neq 0$ the Pad\'e motivated N$^3$LO results for
$\Lambda_{\overline{MS}}^{(4)}$ and $\alpha_s(M_Z)$ turned out
to be almost identical to the NNLO ones (provided the statistical
error bars are taken into account, see Table 1). Moreover, the difference
$\Delta^{NNLO}$=
$|(\Lambda_{\overline{MS}}^{(4)})^{N^3LO}-
(\Lambda_{\overline{MS}}^{(4)})^{NNLO}|$,
which can be considered
as the measure of the theoretical uncertainties of the NNLO
results, is drastically smaller then the NLO correction term
$\Delta^{NLO}$=
$|(\Lambda_{\overline{MS}}^{(4)})^{NNLO}-
(\Lambda_{\overline{MS}}^{(4)})^{NLO}|$. The similar tendency
$\Delta^{NNLO}<<\Delta^{NLO}$ is taking place in the case
of the fits without twist-4 corrections. These observed properties
indicate the reduction of the theoretical errors due to cutting
the analyzed perturbative series at the different orders.

It is known that the inclusion of the higher-order perturbative
QCD corrections into the  comparison with the
experimental data is decreasing the scale-scheme theoretical
errors of the results for $\Lambda_{\overline{MS}}$ and thus
$\alpha_s(M_Z)$ (see e.g. Refs.\cite{ChK,EGKS}). Among the ways of
probing the scale-scheme uncertainties are the scheme-invariant
methods, namely the principle of minimal sensitivity \cite{PMS}
and the effective charges approach \cite{Gr}, which is known to be
identical to the  scheme-invariant perturbation theory,
developed in Refs.\cite{DG,GShT,BL}. These methods can be used to
estimate unknown N$^3$LO corrections to the definite physical
quantities \cite{KSt}.
Note that the estimates of Ref.\cite{KSt} are in agreement
with the results of applications of the Pad\'e resummation technique
(see Ref.\cite{SEK}). Therefore, we can conclude that the application
of the methods of the Pad\'e approximants is leading to the reduction
of the scale-scheme dependence uncertainties of the values of
$\alpha_s(M_Z)$. We are presenting now the 
outcomes of the fits of the CCFR'97 experimental data
for the $xF_3$ SF, obtained with considering twist-4 parameters
as the additional free parameters of the fit:
\begin{eqnarray}
NLO~HT~free~\alpha_s(M_Z) =0.125^{+0.008}_{-0.010}(stat)
\\ \nonumber
\pm 0.005(syst)
\pm 0.009(theory)
\end{eqnarray}
\begin{eqnarray}
NNLO~HT~free~\alpha_s(M_Z)=0.117^{+0.006}_{-0.007}(stat)
\\ \nonumber
\pm 0.005(syst)
\pm 0.002(theory)
\end{eqnarray}
where the systematic uncertainties are taken from the experimental
analysis of Ref.\cite{CCFR97} and the theoretical uncertainties in the results
of Eq.(12) [eq.(13)] are estimated by the diffferences between the
central values of the outcomes of the NNLO and NLO [N$^3$LO and NNLO] fits,
presented in Table 1, plus the arbitrariness in the application of
the $\overline{MS}$-scheme
matching condition of Ref.\cite{ChKS}, which following the
considerations of Ref.\cite{ShSM} we estimate as $\Delta\alpha_s(M_Z)=
\pm 0.001$.

It can be seen that due to the large overall number of the fitted
parameters the results of Eqs.(12),(13) have rather large statistical
uncertainties. It is possible to decrease their values by fixing the
concrete form of the twits-4 parameter $h(x)$. For example, parametrizing
$h(x)$ through the infrared renormalon model of Ref.\cite{DW}, supported
by the independent considerations of Ref.\cite{AZ}, we arrive to the following
values of $\alpha_s(M_Z)$, extracted in Ref.\cite{KKPS2} from the
CCFR'97 data for $xF_3$ SF:
\begin{eqnarray}
NLO~HT~of \cite{DW}~\alpha_s(M_Z) =0.121 \pm 0.002(stat)
\\ \nonumber
\pm 0.005(syst)
\pm 0.006(theory)
\end{eqnarray}
\begin{eqnarray}
NNLO~HT~of \cite{DW}~\alpha_s(M_Z)=0.117\pm 0.002(stat)
\\ \nonumber
\pm 0.005(syst)
\pm 0.003(theory)~.
\end{eqnarray}
Note, however, that we did not yet check the validity of the estimates
of the theoretical error bars of the result of Eq.(15) with the help
of the method of the [0/2] Pad\'e approximant. We are planning to perform this
analysis in another work.

{\bf Acknowledgments}

One of us (ALK) is grateful to S.~Narison for the hospitality
in Montpellier during the very productive QCD-97 Workshop.
This work is supported by the Russian Fund for Fundamental Research,
Grant N 96-02-18897. The work of G.P. was supported by CICYT
(Grant N AEN96-1773) and Xunta de Galicia (Grant N XUGA-20604A96).

\end{document}